\documentclass[runningheads]{llncs}
\usepackage[T1]{fontenc}
\usepackage{graphicx}
\usepackage[english]{babel}
\usepackage{tabulary}
\usepackage{algorithm, algcompatible}
\usepackage{algpseudocode}
\usepackage{bm}

\def\*#1{\mathbf{#1}}
\def\+#1{\mathcal{#1}}
\def\-#1{\mathbb{#1}}
\def\~#1{\mathrm{#1}}

\usepackage{color}
\usepackage{hyperref}

\begin{document}
\title{Multimodal 3D Brain Tumor Segmentation\\with Adversarial Training and\\Conditional Random Field}
\titlerunning{Multimodal 3D Brain Tumor Segmentation}
\author{Lan Jiang \and
Yuchao Zheng \and
Miao Yu \and
Haiqing Zhang \and
Fatemah Aladwani \and
Alessandro Perelli\orcidID{0000-0002-0511-2293}}
\authorrunning{Lan Jiang et al.}
\institute{Centre for Medical Engineering and Technology, University of Dundee, UK\\
\email{\{2402177, 2402169, 2402087, 2402056, 2429072, aperelli001\}@dundee.ac.uk}}
%

\maketitle              

\vspace{.9cm}

\begin{abstract}
Accurate brain tumor segmentation remains a challenging task due to structural complexity and great individual differences of gliomas. Leveraging the pre-eminent detail resilience of CRF and spatial feature extraction capacity of V-net, we propose a multimodal 3D Volume Generative Adversarial Network (3D-vGAN) for precise segmentation. The model utilizes Pseudo-3D for V-net improvement, adds conditional random field after generator and use original image as supplemental guidance. Results, using the BraTS-2018 \cite{menze2014} dataset, show that 3D-vGAN outperforms classical segmentation models, including U-net, Gan, FCN and 3D V-net, reaching specificity over 99.8\%. 

\keywords{Multimodal Segmentation  \and Generative Adversarial Network \and Brain tumor.}
\end{abstract}

\section{Introduction}
Multi-modal Magnetic resonance imaging (MRI) plays an important role nowadays, which is an imaging technique that can produce high-quality images of the human body. It provides a wealth of information for clinical diagnosis and biomedical research, especially the anatomical structure of the brain \cite{zhang2011}. At the same time, various kinds of brain tumors have different performances in MRI data. Based on these, the automatic and accurate classification of MRI images greatly magnifies the diagnostic value of MRI. 

Several segmentation methods have been applied to MRI of brain tumors, containing a combination of traditional and deep learning networks. Different MRI modality has its own specific pathological features. FLAIR images provide more complete information about the tumor boundary for high-intensity signals in edematous areas. Whereas T1 images can reflect clear differences between brain tumors and other healthy tissue. T1c images use low and high-intensity features, representing non-enhanced areas while T2 images play an important role in differentiating healthy brain tissue and can be used to distinguish between diseases caused by cytotoxic spread or extensive edema \cite{ali2020}, \cite{fang2022}.

\subsection{Related Works}

Generative adversarial network (GAN) includes generators and discriminators which can be used for image segmentation, also popular in brain segmentation \cite{neelima2022}. V-net, the 3D U-net, is also a network commonly used in segmentation. 

Recently, there are many studies have been founded based on GANs for MRI brain tumor segmentation. Among these, many kinds of GANs are used to segment MRI brain tumors which also combine some classical networks like V-net, which are shown in Table \ref{tab:Intro}.

\begin{table}
\caption{Common methods for image segmentation and application of GAN.}\label{tab:Intro}
{\renewcommand{\arraystretch}{1.1} 
\begin{tabular}{>{\centering\arraybackslash}p{.3\textwidth}|>{\centering\arraybackslash}p{.2\textwidth}|>{\centering\arraybackslash}p{.3\textwidth}|c}
\hline
\bf Segmentation Method &  \bf Dataset & \bf Performance & \bf GAN\\
\hline
MFF-DNet &  BraTS-2015 & Precision (whole \& core tumor) 0.92 \& 0.90 & - \\
\hline
Optimal
DeepMRSeg \cite{neelima2022} &  BraTS-2018 & Accuracy: 0.914, 0.917 & - \\ \hline
DL-AHS (SHPT-Net, RESU-Net) \cite{alrashedy2022} &  ADNI,
NITRIC & Accuracy: 97\%, 94.34\% & DC-GAN \\ \hline
RUDA \cite{havaei2017} &  MRBrains18 & Accuracy: 93.80\% & - \\ \hline
RescueNet \cite{nema2020} &  BraTS-2015, BraTS-2017 & DICE: 0.9401\%, 0.9463\% & CycleGAN \\ \hline
SSimDCL \cite{guven2023} &  Data Brain -MRI & DICE: 0.9940 & DCLGAN \\ \hline
3D AGSE-VNet \cite{guan2022} &  BraTS-2020 & Dice (whole, core, enhanced): 0.68, 0.85, 0.70 & - \\ \hline
DualMMP-GAN, CACNN-WNet \cite{zhu2022} &  BraTS-2018 & Dice: 0.830 $\pm$
0.154 & CycleGAN \\ \hline
Vox2Vox \cite{cirillo2021} &  BraTS-2020 & Dice: 93.40\%, 92.49\%, 86.48\% & 3D-GAN \\
\hline
\end{tabular}}
\end{table}

Depending on the type and format of the medical image, the segmentation of the image can be subdivided into 2D segmentation and 3D segmentation. The analysis shows that 2D segmentation and 3D segmentation have their own features. The main difference between 2D and 3D image segmentation is in the processing elements. For 2D segmentation, the resulting segmentation can contain inconsistencies, no surface, and lost important 3D contexts. Therefore, the development of a 3D segmentation method is desired for more accuracy. Therefore, based on the above background, we propose a new MRI brain tumor segmentation network that can achieve 3D segmentation of tumors with good results by considering classical networks such as V-net and GAN. 

We consider three tumor subregions for brain segmentation: whole tumor (WT), enhance tumor (ET), and tumor core (TC).

\subsection{Main Contribution}

The proposed framework includes the following contributions: 
\begin{itemize}
\item We propose a novel 3D-vGAN model for MRI image segmentation which alleviates the problem of the rough boundary of segmentation results due to the complex and variable shape of brain tumors in the absence of manually annotated data. 
\item Combine Conditional random fields (CRF) to recover details and improve network performance. We input the original image to the discriminator as an additional informative guidance to further improve the network performance. 
\end{itemize}

The paper is organized as follow: Section \ref{sec:3DvGAN} presents the proposed segmentation framework called 3D-vGAN and in Section \ref{sec:res} we summarize the results on the MRI clinical dataset BraTS-2018 \cite{menze2014}.

\section{3D-vGan Network pipeline}\label{sec:3DvGAN}

\begin{figure}[!h]
    \centering
    \includegraphics[width=\textwidth]{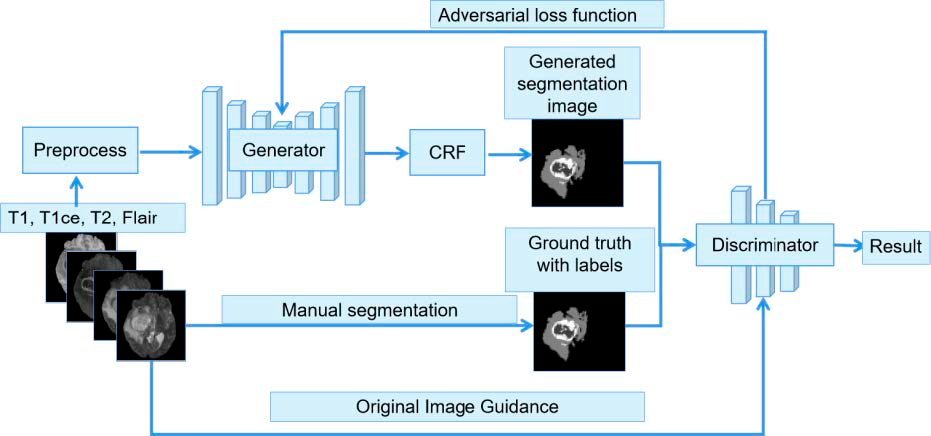}
    \caption{3D-vGAN overall network architecture.}
    \label{fig:Meth2}
\end{figure}

Inspired by 3D GAN network segmentation \cite{radford2015}, the basic model is based on the above DCGAN network, and four different modes of brain tumor MRI images are selected for data input. The generator part is composed of classical V-Net segmentation network and conditional random field for image segmentation. The discriminator is composed of multi-layer CNN, which is used to give the identification results, and feedback the generator through the adversarial loss function to improve the generator generation capability. In addition, the original image is added as an additional information input for guidance to improve the discriminator's identification ability. The overall network structure of this study is shown in Fig. \ref{fig:Meth2}.

\subsubsection{The structure of Generation and Discriminator}
\begin{figure}[!h]
    \centering
    \includegraphics[width=\textwidth]{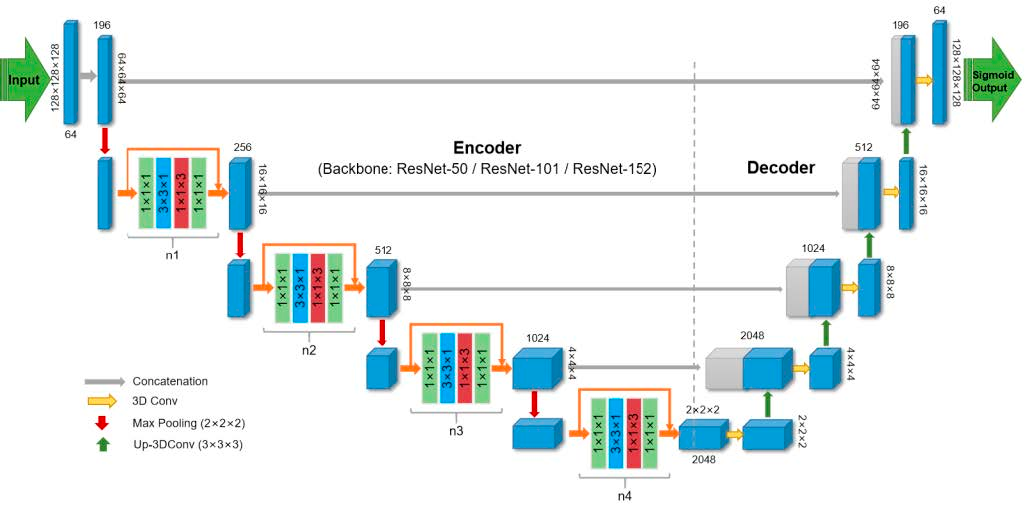}
    \caption{The model structure of the Generator.}
    \label{fig:Meth3}
\end{figure}

The 3D-vGAN model is based on the architecture of GAN \cite{goodfellow2020}, which includes a generator and discriminator. The generator, illustrated by Fig. \ref{fig:Meth3}, is constructed by V-Net \cite{cciccek2016} with Residual Nets \cite{he2016} as backbone.

V-Net is the 3D version of U-Net \cite{ronneberger2015} with similar encoder-decoder structure. The repeated number $n_i\, (i = 1, 2, 3, 4)$ of residual blocks in encoder is determined by the layer structure of different backbone, including ResNet-50, ResNet-101, and ResNet-152, as shown in Table \ref{tab:3D-GAN}.

\begin{table}[!h]
\caption{The setting of $n_i\, (i = 1, 2, 3, 4)$}\label{tab:3D-GAN}
{\renewcommand{\arraystretch}{1.1} 
\begin{tabular}{>{\centering\arraybackslash}p{.3\textwidth}|
>{\centering\arraybackslash}p{.15\textwidth}|
>{\centering\arraybackslash}p{.15\textwidth}|
>{\centering\arraybackslash}p{.15\textwidth}|
>{\centering\arraybackslash}p{.15\textwidth}}
\hline
\bf Backbone & \boldmath $n_1$ & \boldmath $n_2$ & \boldmath $n_3$ & \boldmath $n_4$\\
\hline
ResNet-50 &  3 & 4 & 6 & 3 \\ \hline
ResNet-101 & 3 & 4 & 23 & 3 \\ \hline
ResNet-152 & 3 & 8 & 36 & 3 \\ \hline
\end{tabular}}
\end{table}

The encoder consists of four down-sampling parts. Each part is constructed from several residual blocks using convolutional layers, followed by instance normalization and Leaky ReLU activation function sequentially. ReLU sets all negative values to zero. On the contrary, Leaky ReLU is to give all negative values a non-zero slope, which is specially designed to solve the Dead ReLU problem \cite{maas2013}. The next part is decoder, which intend to concatenate the previous information with the corresponding deconvolutional layers. The decoder contains four up-sampling parts. Each part is built from a 3D transposed convolution with kernel size 3, stride 2, and same padding. After that, the instance normalization and Leaky ReLU activation $(\mathrm{alpha} = 0.2)$ function are also performed.

\begin{figure}[!h]
	\centering
	\includegraphics[width=\textwidth]{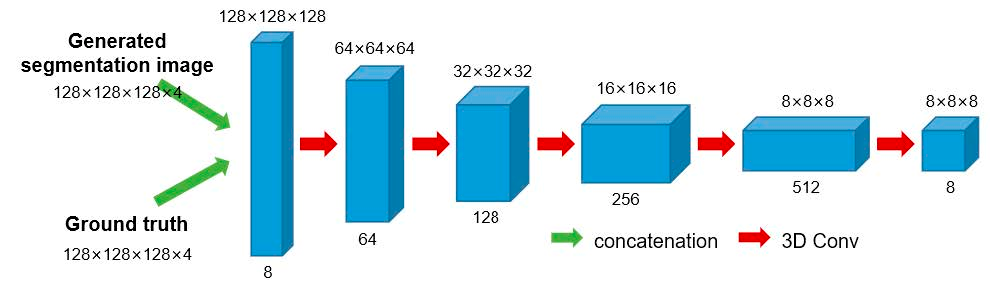}
	\caption{The model structure of the Discriminator \cite{cirillo2021}.}
	\label{fig:Meth5}
\end{figure}

The reason why we choose 3D convolution is that voxels may contain more valuable structural information than 2D pixels, which will definitely promote the segmentation ability of the networks. However, this brings challenge to computer’s calculation capacity and the running speed of training process. In order to reduce the parameters, increase the non-linearity, and improve the fitting ability, Pseudo-3D \cite{qiu2017} instead of real 3D convolution is applied in each Residual blocks. In Pseudo-3D blocks \cite{qiu2017}, the 3D (width, height, depth) dimension can be grouped into a 2D slice dimension (S) and a 1D depth dimension (D). Instead of using a complete 3D convolution, we consider a stacked structure by making 2D filter (S) followed by 1D filter (D) in a cascaded manner, which is given by
\begin{equation}
	(I + DS) \*x_t := \*x_t + D(S(\*x_t)) = \*x_{t+1}
\end{equation}
where $\*x_t$ and $\*x_{t+1}$ represent the input and output of each unit and $F$ is a non-linear residual function. The topology of the discriminator is conventional, which is used to compare the generated segmentation image with the ground truth. It is formed by stacking multiple 3D convolutions, and each convolution is set by kernel size 4, stride 1 and same padding, as shown in Fig. \ref{fig:Meth5}.

\subsubsection{Loss Function}\label{sec:Loss}

The loss function includes two parts, namely the loss function of module $G$ and the loss function of module $D$. Here we define, the original image is $x$, the manually segmented ground texture is $\*y$, and the segmented predicted image generated by the generator is $\hat{\*y}_0$. The loss function generated includes two parts: (1) When the tensor is $1$, the $L_2$ loss between $\*x$ and output $\hat{\*y}$ from the discriminator. (2) The product of scalar weight coefficient $\alpha, \;(\alpha \geq 0)$ and the generalized dice loss between $\*y$ and $\hat{\*y}$. It can be expressed as:
\begin{equation}
	L_G = L_2 [D(\*x, \hat{\*y}), 1] + \alpha GDL(\*y, \hat{\*y})
\end{equation}
where GDL is the Generalized Dice Loss. The discriminator loss function also includes two parts: (1) When the tensor is $1$, the $L_2$ loss between $\*x$ and output $\*y$ from the discriminator. (2) When the tensor is $0$, the $L_2$ loss between $x$ and output $\hat{y}$ from the discriminator. It can be expressed as:
\begin{equation}
	L_D = L_2 [D(\*x, \*y), 1] + L_2[D(\*x, \hat{\*y}), 0]
\end{equation}

From the above, when $\alpha = 0$, the network behaves as a pure GAN network, and the loss function only needs to minimize the unsupervised loss given by the discriminator; And when $\alpha$ approaching infinity, the network ignores the role of discriminator, and finally behaves as a 3D V-net. Therefore, how to obtain accurate segmentation results through confrontation training requires $\alpha$ size.

\subsubsection{Condition Random field Module}

Conditional random field is a conditional probability distribution model, specifically as follows:
\begin{equation}
	P(Y_v| X, Y_w, w \neq v) = P(Y_v| X, Y_w, w \sim v)
\end{equation}

\begin{figure}[!h]
	\centering
	\includegraphics[width=.8\textwidth]{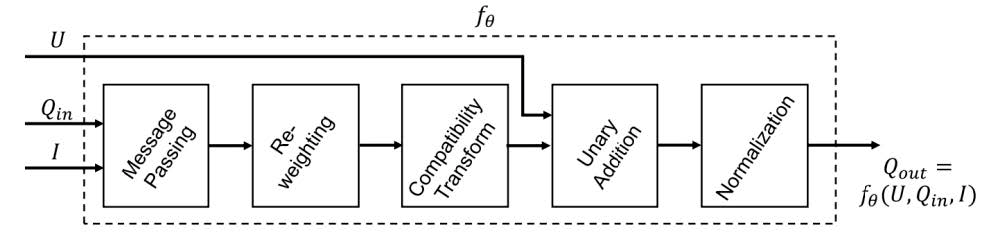}
	\caption{Flow Chart of Conditional Random Fields.}
	\label{fig:Meth6}
\end{figure}

Based on this, \cite{zheng2015} proposed CRF-RNN, that is the conditional random field with Gaussian binary potential function and mean value approximate reasoning is formulated as a recurrent neural network. Specifically, each iteration of the algorithm includes five steps, shown in Fig. \ref{fig:Meth6}: message passing, re-weighting, compatibility transform, unary addition, and normalization. As shown in Fig. \ref{fig:Meth6}, each step of the iterative process is programmed as a sub layer, and all sub layers are superposed and iterative training is conducted to form a conditional random field in the form of a cyclic neural network.

In the initialization step, the operation $Q_i(l) \leftarrow \exp(U_i(l))$, where $Z_i = \sum \exp(U_i(l))$ is performed. $Q_i(l)$ is a simple distribution, $U_i(l)$ is the unary potentials. And this step is the same as applying a soft-max function to all of the labels at each pixel and the unary potentials $U$. Then is the message passing, which applies $M$ Gaussian filters on $Q$ values. The re-weighting step is taking a weighted sum of the $M$ filter outputs form the previous step, for each class label $l$. This can be implemented using convolution with $\alpha$ $1 \times 1$ filter with $M$ input channels, and one output channel. The third step is compatibility transform. The previous output is shared between the labels to a varied extent, depending on the compatibility between these labels. The fourth step is unary addition: the $Q$ is subtracted element-wise from the unary inputs $U$. The final step can be considered as another soft-max operation with no parameters. The overall CRF-RNN algorithm is depicted in Algorithm \ref{table:CRF}.

\begin{algorithm}[!h]
	\caption{CRF-RNN algorithm}
	\label{table:CRF}
	\begin{algorithmic}
		\State $Q_i(l) \leftarrow \exp(U_i(l))$, for all $i$ \Comment{Initialization}
		\While {not converged}
		  \State $\tilde{Q}^{m}_i(l) \leftarrow \sum_{j\neq i} k^m(f_i,f_j) Q_j(l)$, for all $m$   \Comment{Message Passing}
		  \State $\breve{Q}_i(l) \leftarrow \sum_{m} w^m\tilde{Q}^{m}_i(l)$ \Comment{Weighting Filter Output}
		  \State $\hat{Q}_i(l) \leftarrow \sum_{l'\in L} \mu(l,l')\breve{Q}_i(l')$ \Comment{Compatibility Transform}
		  \State $\breve{Q}_i(l) = U_i(l) - \hat{Q}_i(l)$\Comment{Adding Unary Potentials}
		  \State $Q_i \leftarrow \frac{1}{Z_i}\exp(\breve{Q}_i(l))$\Comment{Normalizing}
		\EndWhile
	\end{algorithmic}
\end{algorithm}

CRF-RNN combines the advantages of CNN and CRF which is mainly used to enhance the spatial continuity of output labels and accurately locate the predicted pixels locally, thus helping to produce smooth and accurate image segmentation results. In our framework, we add the CRF-RNN layer, insert it into the generator network, and use the CRF-RNN as the post-processing program of the U-Net to recover the details of the output segmentation result graph, so as to improve the performance of the segmentation network. 

The whole training process of 3D-vGAN is summarized by Algorithm \ref{table:DMDL}. 

\renewcommand{\algorithmicrequire}{\textbf{Input:}}
\renewcommand{\algorithmicensure}{\textbf{Output:}}

\begin{algorithm}[!h]
	\caption{Training procedure of 3D-vGAN}
	\label{table:DMDL}
	\begin{algorithmic}
		\Require 3D brain MRI images $\*x$ and segmented ground truth $\*y$ 
		\Ensure Segmented images $\hat{\*y}$
		\For{$training$ $iterations$}
		\For{$k$ $steps$}
		\State Sample mini-batch of $m$ noise samples $\{z^{(1)}, \ldots, z^{(m)}\}$ from noise prior $p_g(\*z)$
		\State Sample mini-batch of $m$ samples $\{x^{(1)}, \ldots, x^{(m)}\}$ from distribution $p_{data}(\*x)$
		\State Update the discriminator by ascending the loss: 
		\State \begin{center} $L_D = L_2[D(\*x, \*y),1] + L_2[D(\*x,\hat{\*y},0)]$ \end{center}
		\EndFor
		\State Sample mini-batch of $m$ noise samples $\{z^{(1)}, \ldots, z^{(m)}\}$ from noise prior $p_g(\*z)$
		\State Update the discriminator by descending the loss: 
		\State \begin{center} $L_G = L_2[D(\*x, \hat{\*y}),1] + \alpha GDL(\*y,\hat{\*y})$ \end{center}
		\EndFor
		\While{$not$ $converged$}
		\State Message Passing
		\State Weighting Filter Outputs
		\State Compatibility Transform
		\State adding Unary Potentials
		\State Normalizing
		\EndWhile
	\end{algorithmic}
\end{algorithm}

First we input the pre-processed images with four dimensions into the generator network and then get the segmentation result. This result is combined with ground truth and then be put into the discriminator, The output of the discriminator will both update the parameters of the generator and discriminator itself. Differ from the general GAN networks, a new loss function is applied to evaluate and update the performance of generator and discriminator. 

\subsubsection{Evaluation Metrics}
To evaluate the performance of the proposed 3D-vGAN network, the following four evaluation metrics were used, and the four metrics are defined as follows. Dice similarity coefficient (DSC) is defined as:
\begin{equation}
	DSC(A,B) = \frac{2\cdot |A \bigcap B|}{|A| + |B|}
\end{equation}
where $A$ represents the ground truth of the brain tumor segmentation and $B$ represents the neural network segmentation of the brain tumor. DSC ranges from zero to one (one is the perfect score). 

The Hausdorff distance is defined as:
\begin{equation}
	H(G,P) = \max\left(\mathrm{supin}_{x\in A, y \in B} d(x,y), \mathrm{supin}_{y \in B, x\in A} d(x,y) \right)
\end{equation}
where $d(x,y)$ denotes the distance of $x$ and $y$, $\mathrm{supin}$ denotes the supremum. This measures how far two subsets of a metric space are from each other. Sensitivity, is also called true positive rate) measures the proportion of actual positives that are correctly identified. Specificity is also called true negative rate measures the proportion of actual negatives that are correctly identified. They are defined as follows:
\begin{equation}
	\mathrm{Sensitivity} = \frac{TP}{TP + FN}\,, \quad\quad \mathrm{Specificity} = \frac{TN}{TN + FP}
\end{equation}
where TP represents true positive, TN represents true negative, FP represents false positive, FN represents false negative.

\section{Results}\label{sec:res}

\subsection{Data Preprocessing}
The dataset is from an open-source competition task, Multimodal Brain tumor Segmentation Challenge 2018 (BraTS-2018) \cite{menze2014}, which contains 100 three-dimensional (3D) images, with 4 MRI modalities (T1, T1c, T2, and FLAIR) per case. The annotations divide the image into three tumor subregions, namely whole tumor (WT), tumor core (TC), and enhancing tumor (ET). Before feeding the data into the neural network, a series of preprocessing operations are performed. Before the training, the following pre-processing shall be performed on the acquired data:
\begin{enumerate}
\item Standardization: Since the input data are multimodal images, which are different from each $255$ other, the $\*z$-score method is used for preprocessing. This method uses the following formula to uniformly
convert the input multimodal data, so as to obtain a specific score and then compare them. Here, $\*x$ represents the value of a single sample, $v$ represents the average of all samples, and $120590$ represents the standard deviation of all samples. $\*z = (\*x - \bm\mu) / \sigma$. 
\item Slice Clipping: Cut the brain tumor data of each mode, and apply patch expansion to change it from the original volume of $240 \times 240 \times 155$. Extract as sub volume $128 \times 128 \times 128$. Slice the modal data, remove the black area on the edge, and discard the disease-free section. Take the four classes to be divided (namely, four regions with labels $0, 1, 2$ and $4$) as the four channels of each target, and combine the standardized and sliced data into four channel data. 
\item Data enhancement: In the case of less training data, over fitting is easy to occur. In this regard, we enhance the existing samples by random 3D rotation, elastic deformation, etc., so that the network has a better generalization effect.
\end{enumerate}

\subsection{Parameter setting}
\begin{figure}[!h]
	\centering
	\includegraphics[width=\textwidth]{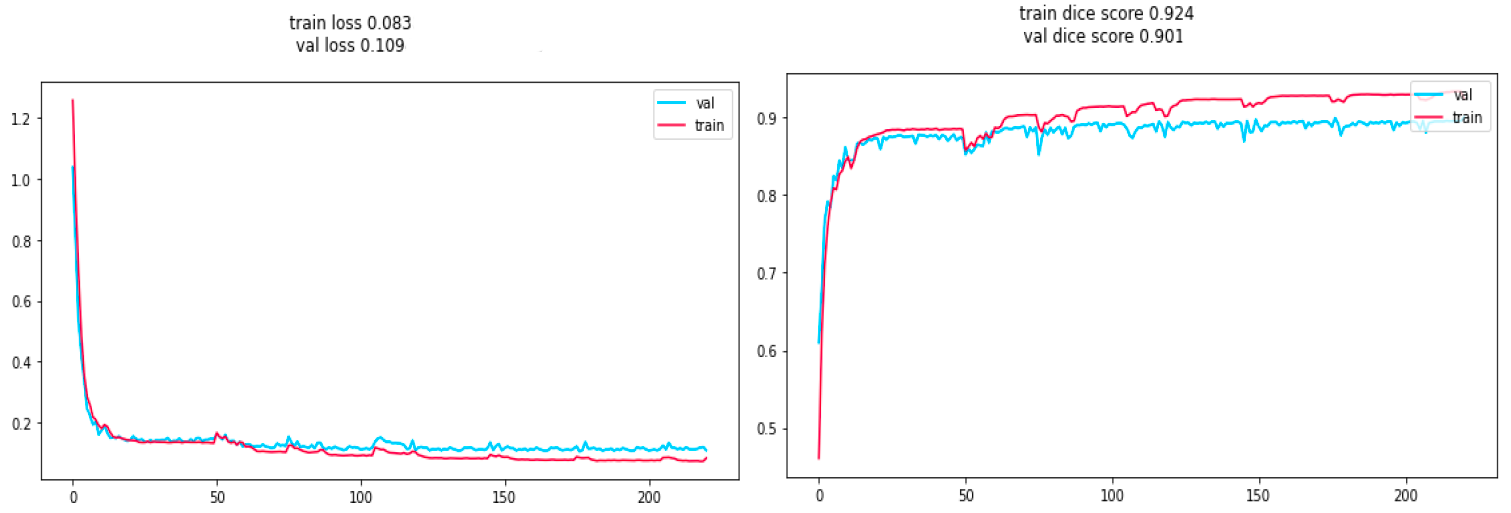}
	\caption{a) Training and validation loss \hspace{1.5cm} b) dice score curves.}
	\label{fig:Res1}
\end{figure}

The data in the dataset is divided according to ten times cross validation. During the experiment, the Adam optimizer is used for training, and the parameters are set to $\lambda = 2 \times 10^{-4}$, $\beta_1 = 0.5$, $\beta_2 = 0.999$. The convolution operation of bottleneck layer in the generator adopts drop regularization with probability of $0.2$. The model is implemented by Python $3.9$, Tensorflow $2.1$ and its Keras library. NVIDIA GeForce RTX 3090 24 GB is used for accelerated operation. The learning rate is set to $0.0001$, and the batch size is $4$, $200$ epochs of network training are conducted. Loss and dice score curves during training and validation are shown in Fig. \ref{fig:Res1}.

\begin{figure}[!h]
	\centering
	\includegraphics[width=\textwidth]{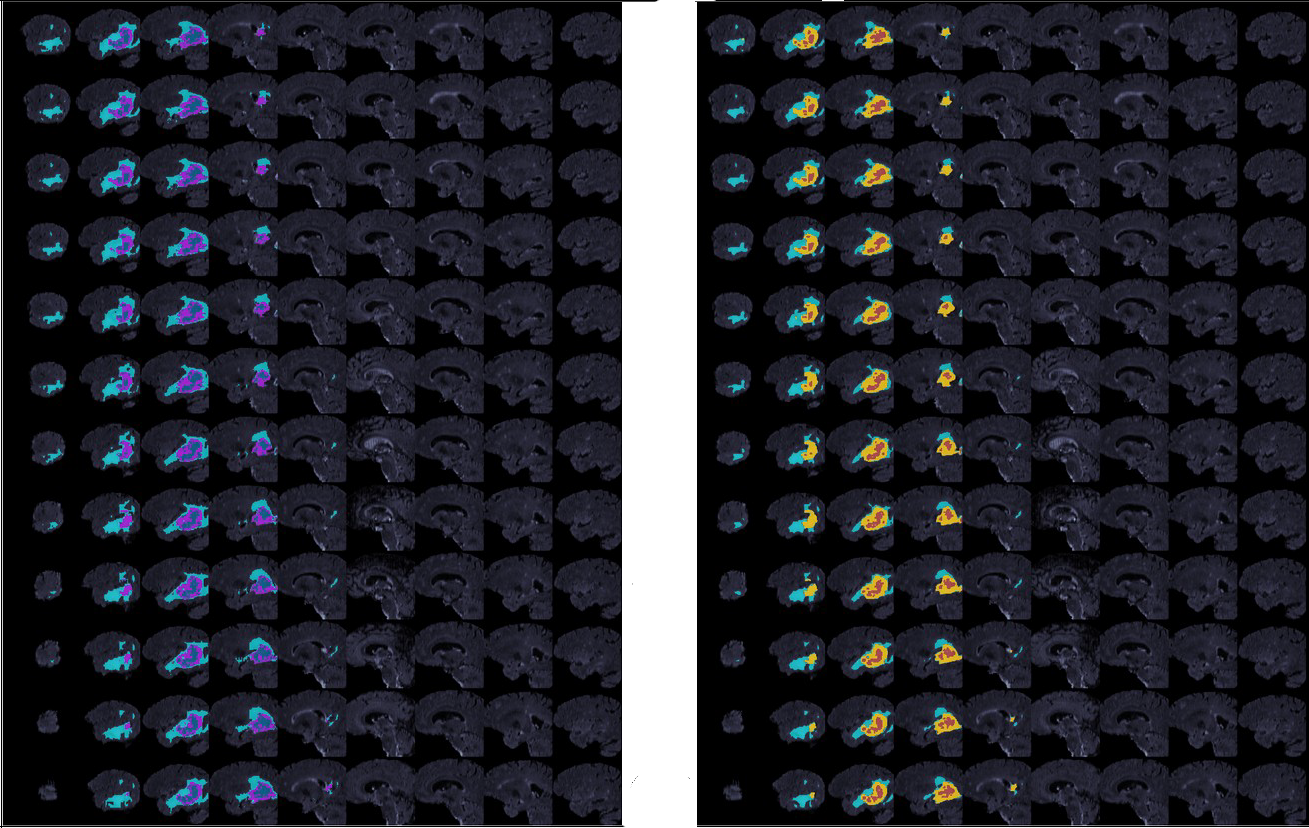}
	\caption{a) Ground truth, \hspace{3cm} b) 3D-vGAN}
	\label{fig:Rec2}
\end{figure}

In Fig. \ref{fig:Res1}(b) after 200 iterations, the dice scores on both the training and validation sets exceeded 0.90, and the visualization results obtained also showed that several parts of the brain tumor were well segmented, especially the whole tumor part. Fig. \ref{fig:Rec2} shows the comparison between the ground truth segmentation label and the qualitative results of the segmented tumors using 3D-vGAN.

\begin{figure}[!h]
	\centering
	\includegraphics[width=\textwidth]{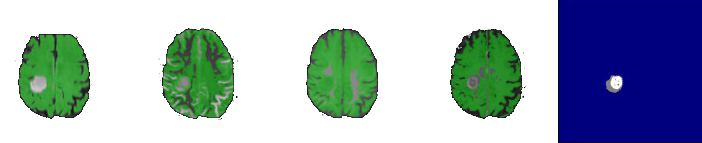}
	\caption{a) T1 \hspace{.9cm} b) T1c \hspace{1.2cm} c) T2 \hspace{.6cm} d) FLAIR images \hspace{.5cm} e) 3D-vGAN}
	\label{fig:Rec1}
\end{figure}

Fig. \ref{fig:Rec1} shows the T1, T1c, T2, and FLAIR input images on a brain sample and the qualitative segmented mask obtained using 3D-vGAN.

In Table \ref{tab:resul_comp} it is shown how our proposed network outperforms the conventional segmentation networks in terms of DSC and specificity, and significantly outperforms in terms of sensitivity. In other conventional networks, the sensitivity is below 0.8, our proposed network has a sensitivity of 0.84; finally also the Hausdorff distance metric is noticeably reduced compared to the other networks. 

\begin{table}[!h]
	\caption{Comparison of 3D-vGAN with other segmentation networks}\label{tab:resul_comp}
	\begin{center}
		\begin{tabular}{ |m{2cm}|c|c|c|m{2cm}| } 
			\hline
			& \bf DSC (\%) & \bf Sensitivity (\%) & \bf Specificity (\%) & \bf Hausdorff Distance (mm) \\ \hline\hline
			\center{\bf U-net} & 74.13 & 78.63 & 95.79 & 26.31 \\ \hline
			\center{\bf GAN}& 67.12 & 72.64 & 65.68 & 34.20 \\\hline
			\center{\bf FCN} & 61.55 & 71.24 & 63.83 & 38.50 \\\hline
			\center{\bf 3D V-net} & 68.83 & 71.84 & 95.79 & 26.31 \\\hline
			\center{\bf 3D v-GAN (ours)} & \bf 82.13 & \bf 84.42 & \bf 99.97 & \bf 11.89 \\
			\hline
		\end{tabular}
	\end{center}
\end{table}

\subsection{Parameters Experiments}

The loss proposed is dependent on the parameter $\alpha$ and the size on network structure is discussed in subsection \ref{sec:Loss}. We evaluated how to adjust the size of $\alpha$ and to check how the discriminator and generator are converging to the  min-max solution. We perform the segmentation on the BraTS2018 dataset for different $\alpha$. The differences are reported in Table \ref{tab:params}; it can be clearly seen when $\alpha = 5$ the performance of the model reached the optimal values. In addition, it can be seen that when discriminator's guiding role increases ($\alpha$ increases), the model achieve better results.

\begin{table}[!h]
	\caption{Parameters evaluation}\label{tab:params}
	\begin{tabular}{ |c|c|c|c|c|c|c|c|c|c|c|c|c| } 
		\hline
		\bf $\alpha$ & \multicolumn{3}{c|}{\bf DSC (\%)} & \multicolumn{3}{c|}{\bf Sensitivity (\%)} & \multicolumn{3}{c|}{\bf Specificity (\%)} & \multicolumn{3}{m{.21\textwidth}|}{\bf Hausdorff Distance (mm)} \\ \hline
		  & ET &  WT & TC & ET &  WT & TC & ET &  WT & TC & ET &  WT & TC \\ \hline\hline
		0 & 37.21 & 56.78 & 24.51 & 44.91 & 63.21 & 47.11 & 98.61 & 98.71 & 99.01 & 73.20 & 76.55 & 99.89 \\ \hline
		1 & 62.10 & 85.41 & 71.35 & 59.13 & 71.88 & 63.17 & 99.37
		& 99.36 & 99.71 & 41.09 & 13.76 & 18.12 \\ \hline
		\bf 5 & \bf 82.67 & \bf 92.15 & \bf 90.97 & \bf 82.11 & \bf 92.16 & \bf 91.03 & \bf 99.97 & \bf 99.81 & \bf 99.86 & \bf 28.99 & \bf 3.42 & \bf 3.19 \\ \hline
		10 & 71.44 & 88.63 & 81.20 & 75.66 & 82.79 & 76.61 & 99.10 & 99.36 & 99.21 & 34.19 & 12.99 & 10.71 \\ \hline
		25 & 61.96 & 87.51 & 75.31 & 60.94 & 77.11 & 46.16 & 99.01 & 98.90 & 99.11 & 30.23 & 9.12 & 7.89 \\
		\hline
	\end{tabular}
\end{table}

\section{Conclusion}

In this paper, we present a brain tumor segmentation method based on an improved generative adversarial network approach, implementing a deep learning network combining conditional random fields and multiple tasks. And by using the multi task learning method, we can automatically segment and enhance the brain tumor image when the details of the brain tumor image are blurred and the contrast is low by designing the mode of shared parameters and concurrently carrying out the two tasks of brain tumor segmentation and image enhancement. It not only provides a new research idea and solution for the segmentation of brain tumors, but also provides a deep learning network based on a generative adversarial approach for the segmentation of regions of interest in MRI multimodal images.

\bibliographystyle{splncs04}
\bibliography{biblio}

\end{document}